\documentclass[copyright,creativecommons]{eptcs}
\providecommand{\event}{ICE'09} 
\providecommand{\volume}{??}
\providecommand{\anno}{2009}
\providecommand{\firstpage}{1}
\providecommand{\eid}{??}

\usepackage{graphicx}

\DeclareGraphicsExtensions{.pdf,.jpeg,.png,.eps}
\usepackage[english]{babel}
\usepackage{dsfont}
\usepackage{latexsym}
\usepackage{amssymb}
\usepackage{amsfonts}
\usepackage{fancybox}
\usepackage{txfonts}
\usepackage{pxfonts}
\usepackage{wasysym}
\usepackage{epsfig}
\usepackage{fancyhdr}
\usepackage[tight,figbotcap,tabbotcap,]{subfigure}
\usepackage[all]{xy}
\usepackage{floatflt}
\usepackage{mypaper}
\usepackage{cmll}
\usepackage{proof}
\usepackage{esvect}

\input xy
\xyoption{all}

\usepackage{url}

\newcommand{\escmathcom}[3]{ \newcommand{#1}[#2]{\mbox{#3}}}
\escmathcom{\mcal}{1}{$\mathcal{#1}$}
\escmathcom{\msf}{1}{\sf{#1}}
\escmathcom{\lrelll}{1}{\xleftarrow{#1}}

\escmathcom{\dynex}{0}{$\hat{\exists}$}
\escmathcom{\dynall}{0}{$\hat{\forall}$}
\escmathcom{\lnlambda}{0}{$\hat{\lambda}$}
\escmathcom{\lnLambda}{0}{$\hat{\Lambda}$}
\escmathcom{\caret}{0}{\^{}}
\escmathcom{\lsem}{0}{$\llbracket$}
\escmathcom{\rsem}{0}{$\rrbracket$}
\escmathcom{\lnvareps}{0}{$\hat{\varepsilon}$}
\escmathcom{\One}{0}{$\mathbf{1}$}
\escmathcom{\Kappa}{0}{$\Box$}
\escmathcom{\Alpha}{0}{$\Theta$}
\escmathcom{\upnu}{0}{$\nu$}
\escmathcom{\lequiv}{0}{$\hat{\equiv}$}
\escmathcom{\downmapsto}{0}{$\downharpoonright$}
\escmathcom{\plus}{0}{$\&$}
\escmathcom{\defeq}{0}{$\ =_{df} \ $}
\escmathcom{\ineg}{0}{$\neg$}
\escmathcom{\grax}{0}{$\vDash$}

\escmathcom{\tranred}{0}{$ \ \rightarrow_{tr} \ $}
\escmathcom{\instred}{0}{$ \ \rightarrow_{ir} \ $}
\escmathcom{\evalred}{0}{$ \ \rightarrow_{er} \ $}
\escmathcom{\tranext}{0}{$ \ \rightarrow_{te} \ $}
\escmathcom{\instext}{0}{$ \ \rightarrow_{ie} \ $}
\escmathcom{\somered}{0}{$ \ \twoheadrightarrow_{r} \ $}
\escmathcom{\patred}{0}{$ \ \rightarrow_{ptr} \ $}
\escmathcom{\paired}{0}{$ \ \rightarrow_{pir} \ $}

\escmathcom{\letexp}{3}{$\msf{let} \ {#1}={#2} \ \msf{in} \ {#3}$}
\escmathcom{\letex}{3}{$\msf{let} \ {#1}={#2} \ \msf{in} \ {#3}$}

\newcounter{observation}
\renewcommand{\theobservation}
{\arabic{observation}}

\newenvironment{obsA}
{\refstepcounter{observation}%
 \begin{description}%
 \item[\textbf{Obs.}
       \theobservation]}%
{\end{description}}

\newcounter{prop}
\renewcommand{\theprop}
{\arabic{prop}}

\newenvironment{propA}
{\refstepcounter{prop}%
 \begin{description}%
 \item[\textbf{Prop.}
       \theprop]}%
{\end{description}}

\title{Towards an Embedding of Graph Transformation in Intuitionistic
  Linear Logic} 
\author{Paolo Torrini \& Reiko Heckel 
\institute{Department of Computer Science, University of Leicester}
\email{\{pt95,reiko\}@mcs.le.ac.uk}
}

\def\titlerunning{Graph Transformation in ILL}
\def\authorrunning{P. Torrini \& R. Heckel}

\begin{document}
\maketitle

\begin{abstract}
  Linear logics have been shown to be able to embed both
  rewriting-based approaches and process calculi in a single,
  declarative framework. In this paper we are exploring the embedding
  of double-pushout graph transformations into quantified linear
  logic, leading to a Curry-Howard style isomorphism between graphs /
  transformations and formulas / proof terms. With linear implication
  representing rules and reachability of graphs, and the tensor
  modelling parallel composition of graphs / transformations, we
  obtain a language able to encode graph transformation systems and
  their computations as well as reason about their properties.
\end{abstract}


\sloppy


\section{Introduction}

Graphs are among the simplest and most universal models for a variety
of systems, not just in computer science, but throughout engineering
and life sciences. When systems evolve, we are generally interested in
the way they change, to predict, support, or react to evolution. Graph
transformation systems (GTS) combine the idea of graphs, as a
universal modelling paradigm, with a rule-based approach to specify
the evolution of systems. The double-pushout approach (DPO)
\cite{EEPT06} is arguably the most mature of the
mathematically-founded approaches to graph transformation, with a rich
theory of concurrency comparable to (and inspired by) those of
place-transition Petri nets and term rewriting systems.

The fact that graph transformations are specified at the level of
visual rules is very important at the intuitive level. However, these
specifications are still operational rather than declarative. In order
to reason about them, and to prove their properties at a
realisation-independent level, a logics-based representation is
desirable. Intuitionistic linear logic (ILL) allows us to reason about
concurrent processes at a level of abstraction which can vary from
statements on individual steps to the overall effect of a longer
computation.  Unlike operational formalisms, linear logics are not
bound to any particular programming or modelling paradigm and thus
have a potential for integrating and comparing different such
paradigms through embeddings \cite{Girard87,abram93}.

What makes ILL well applicable to GTS is the handling of resources and
the way this allows for expressing creation/deletion of graph
components. However, expressing the notion of pattern matching used in
DPO in logic terms is not straightforward --- to this purpose we
extend ILL with a form of resource-bound quantification.  In this
paper we propose an embedding of DPO-GTS in a variant of quantified
intuitionistic linear logic with proof terms (QILL).  Our translation
relies on a preliminary algebraic presentation of DPO-GTS in terms of
an SHR-style formalism \cite{Ferrari06}, which gives us syntactic
notions of graph expression and transformation rule.


QILL is based on linear $\lambda$-calculus \cite{BeEA93,CePf02,Pf02},
and is obtained by adding to ILL standard universal quantification
($\forall$), and a form of resource-bound existential quantification
($\dynex$), associating a linear resource to each variable --- in this
respect quite different from the intensional quantifiers in
\cite{pym02}. In order to deal with the nominal aspect, we use
non-quantifiable constants, treated as linear resources, to which
individual variables may refer --- unlike nominal logic
\cite{Pitts01,CaCa04}, where names can be treated as bindable atoms.

We translate algebraic graph expressions to linear $\lambda$-calculus,
so that component identity is represented in the proof-terms, whereas
typing information and connectivity is represented in the logic
formula. We obtain a Curry-Howard style isomorphism between graph
expressions and a subset of typing derivations, and between graphs and
a subset of logic formulas (graph formulas) modulo linear equivalence.
This can be extended to a mapping from GTS runs into typing
derivations, and from reachable graphs into logic formulas. We hope
that this approach will offer the possibility of applying
goal-directed proof-methods \cite{Mil07,Dixon06} to the verification
of well-formedness and reachability properties in GTS.

\section{Basic concepts and intuition}

Here we give a brief introduction of the main concept and the ideas
behind the approach we are working on, before getting further into
details.

\subsection{Hypergraphs and their Transformations}

Graph transformations can be defined on a variety of graph structures,
including simple edge or node labelled graphs,
attributed or typed graphs, etc. In this paper we prefer typed
hypergraphs, their n-ary hyperedges to be presented as
predicates in the logic.

A hypergraph $(V,E,\msf{s})$ consists of a set $V$ of vertices, a set
$E$ of hyperedges and a function $\msf{s}: E \to V^\ast$ assigning
each edge a sequence of vertices in $V$. A morphism of hypergraphs is
a pair of functions $\phi_V: V_1 \to V_2$ and $\phi_E: E_1 \to E_2$
that preserve the assignments of nodes, that is, $\phi_V^\ast \circ
\msf{s}_1 = \msf{s}_2 \circ \phi_E$.

Typed hypergraphs are defined in analogy to typed graphs. Fixing a
type hypergraph $TG = (\mcal{V},\mcal{E},\msf{ar})$ we establish sets
of node types $\mcal{V}$ and edge types $\mcal{E}$ as well as defining
the arity $\msf{ar}(a)$ of each edge type $a \in \mcal{E}$ as a
sequence of node types.  A $TG$-typed hypergraph is a pair $(HG,
type)$ of a hypergraph $HG$ and a morphism $type: HG \to TG$.  A
$TG$-typed hypergraph morphism $f: (HG_1, type_1) \to (HG_2, type_2)$
is a hypergraph morphism $f: HG_1 \to HG_2$ such that $type_2 \circ f
= type_1$.

A \emph{graph transformation rule} is a span of injective hypergraph
morphisms $s = (L \lrel l K \rrel r R)$, called a \emph{rule span}.  A
hypergraph transformation system (GTS) \ $\mcal{G} = \langle TG, P,
\pi, G_0 \rangle$ consists of a type hypergraph $TG$, a set $P$ of
rule names, a function mapping each rule name $p$ to a rule span
$\pi(p)$, and an initial $TG$-typed hypergraph $G_0$.

A \emph{direct transformation} $G \Rrel{p, m} H$ is given by a
\emph{double-pushout (DPO) diagram} as shown below, where (1), (2) are
pushouts and top and bottom are rule spans. If we are not interested
in the match and/or rule of the transformation we will write $G
\Rrel{p} H$ or just $G \Rrel{} H$.

For a GTS $\mcal{G} = \langle TG, P, \pi, G_0 \rangle$, a derivation
$G_0 \Rrel{} G_n$ in $\G$ is a sequence of direct transformations $G_0
\Rrel{r_1} G_1 \Rrel{r_2} \cdots \Rrel{r_n} G_n$ using the rules in
$\G$. The set of all hypergraphs reachable from $G_0$ via derivations
in $\mcal{G}$ is denoted by $\mcal{R}_{\tiny \mcal{G}}$.

\centerline{ \xymatrix{ L  \ar@{}[dr]|{(1)} \ar@{->}[d]_{m} & K
\ar@{}[dr]|{(2)} \ar[l]_{l} \ar[r]^{r} \ar@{->}[d]^{d} &
R  \ar@{->}[d]^{m^*}\\
G  & D  \ar[l]^{g}_{} \ar[r]^{}_{h} & H}}

Intuitively, the left-hand side $L$ contains the structures that must be
present for an application of the rule, the right-hand side $R$ those that are
present afterwards, and the gluing graph $K$ specifies the
``gluing items'', i.e., the objects which are read during
application, but are not consumed.

Operationally speaking, the transformation is performed in two steps.
First, we delete all the elements in $G$ that are in the image of $L
\setminus l(K)$ leading to the left-hand side pushout (1) and the
intermediate graph $D$.  Then, a copy of $L \setminus l(K)$ is added
to $D$, leading to the derived graph $H$ via the pushout (2).

It is important to point out that the first step (deletion) is only
defined if a built-in application condition, the so-called gluing
condition, is satisfied by the match $m$.  This condition, which
characterises the existence of pushout (1) above, is usually presented
in two parts.
\begin{description}
\item [Identification condition:] Elements of $L$ that are meant to be
  deleted are not shared with any other elements, i.e., for all $x \in
  L \setminus l(K)$, $m(x) = m(y)$ implies $x=y$.
\item [Dangling condition:] Nodes that are to be deleted must not be
  connected to edges in $G$, unless they already occur in $L$, i.e.,
  for all $v \in V_G$ such that $v \in m_V(L_V)$, if there exists $e
  \in E_G$ such that $\msf{s}(e) = v_1 \dots v \dots v_n$, then $e \in
  m_E(L_E)$.
\end{description}

The first condition guarantees two intuitively separate properties of
the approach: First, nodes and edges that are deleted by the rule are
treated as resources, i.e., $m$ is injective on $L \setminus l(K)$.
Second, there must not be conflicts between deletion and preservation,
i.e., $m(L \setminus l(K))$ and $m(l(K)$ are disjoint.

The second condition ensures that after the deletion of nodes, the remaining
structure is still a graph and does not contain edges short of a node.
It is the first condition which makes linear logic so attractive for
graph transformation. Crucially, it is also reflected in the notion of
concurrency of the approach, where items that are deleted cannot be
shared between concurrent transformations.

There is a second, more declarative interpretation of the DPO diagram
as defining a rewrite relation over graphs.  Two graphs $G, H$ are in
this relation $G \Rrel{p} H$ iff there exists a morphism $d: K \to D$
from the interface graph of the rule such that $G$ is the pushout
object of square (1) and $H$ that of square (2) in the diagram above.
In our algebraic presentation we will adopt this more declarative
view.

As terms are often considered up to renaming of variables, it is
common to abstract from the identity of nodes and hyperedges
considering hypergraphs up to isomorphism. However, in order to be
able to compose graphs by gluing them along common nodes, these have
to be identifiable. Such potential gluing points are therefore kept as
the \emph{interface} of a hypergraph, a set of nodes $I$ embedded into
$HG$ by a morphism $i: I \to HG$.

An abstract hypergraph $i: I \to [HG]$ is then given by the isomorphism
class $\{i': I \to HG' \mid \exists \mbox{ isomorphism } j: HG \to HG'
\mbox{ such that } j \circ i = i'\}$.

If we restrict ourselves to rules with interfaces that are discrete
(i.e., containing only nodes, but no edges.), a rule can be
represented as a pair of hypergraphs with a shared interface $I$, i.e., $\Lambda I. L
\Rrel{} R$, such that the set of nodes $I$ is a subgraph of both $L, R$.
This restriction does not affect expressivity in describing individual
transformations because edges can be deleted and recreated, but it
reduces the level concurrency.  In particular, concurrent
transformation steps can no longer share edges because only items that
are preserved by both rules can be accessed concurrently.

\subsection{Linear logic}

ILL is a resource-conscious logic that can be obtained from
intuitionistic logic, in terms of sequent calculus, by restricting the
application of standard structural rules \emph{weakening} and
\emph{contraction}. ILL formulas can be interpreted as partial states
and express transitions in terms of consequence relation
\cite{CeSc06}. Tensor product ($\otimes$) can be used to represent
parallel composition, additive conjunction ($\plus$) to represent
non-deterministic choice, and linear implication ($\multimap$) to
express reachability. Unlimited resources can be represented via $!$.

ILL has an algebraic interpretation based on quantales and a
categorical one based on symmetric monoidal closed categories
\cite{BeEA93}, it has interpretations into Petri-nets, and for its
$\lor$-free fragment, it has a comparatively natural Kripke-style
semantics based on a ternary relation \cite{Ishi01} in common with
relevant logics. ILL can be extended with quantifiers. It can also be
enriched with proof terms, thus obtaining \emph{linear
  $\lambda$-calculus} \cite{BeEA93,Pf02}, where linear
$\lambda$-abstraction and linear application require that the
abstraction/application term is used only once.  We are going to rely
on an operational semantics in terms of natural deduction rules,
following \cite{Pf02}.

Proofs can be formalised in terms of natural deduction, based on
introduction/elimination rules closely related to the
constructor/destructor duality in recursive datatypes
\cite{ProofTh00}. Proof normalisation guarantees modularity, meaning
that detours in proofs can be avoided, i.e. one does not need to
introduce a constructor thereafter to eliminate it. Proof
normalisation shows that introducing a constructor brings nothing more
than what it is taken away by eliminating it.


\subsection{GTS in QILL} \label{GTS}


We are going to give a representation of graphs and transformations in
terms of provable sequents. Graphs can be represented by formulas of
form $\dynex \overline{x:A}.  L_1 \ (\overline{x}_1) \otimes \ldots
\otimes L_k \ (\overline{x}_k)$ where $\overline{x:A}$ is a sequence
$x_1:A_1, \ldots, x_j:A_j$ of typed variables and
$\overline{x}_1,\ldots,\overline{x}_k \subseteq \overline{x}$. A DPO
rule (we consider rules with interfaces made only of nodes) can be
represented as $\forall \overline{x:A}. \alpha \multimap \beta$ where
$\alpha,\beta$ are graph expressions. Given rules

$$P_1
= \forall \overline{x}_1. \alpha_1 \multimap \beta_1, \ \ldots, \ P_k =
\forall \overline{x}_k. \alpha_k \multimap \beta_k$$

a sequent $G_0, P_1,\ldots,P_k \Vdash G_1$ can express that graph
$G_1$ is reachable from the initial graph $G_0$ by applying them,
abstracting away from the application order, each occurrence resulting
into a transformation step. A sequent $G_0, !P_1,\ldots,!P_k \Vdash
G_1$ can express that $G_1$ is reachable from $G_0$ by the same rules,
regardless of whether or how many times they must be applied.
The parallel applicability of rules $\forall \overline{x}_1. \alpha_1
\multimap \beta_1$, $\forall \overline{x}_2. \alpha_2 \multimap
\beta_2$ can be represented as applicability of $\forall
\overline{x}_1, \overline{x}_2. \alpha_1 \otimes \alpha_2 \multimap
\beta_1 \otimes \beta_2$.

Logic formulas can be used also to specify graphs according to their
properties --- such as matching certain patterns.  Additive
conjunction ($\plus$) can then be used to express choice, and additive
disjunction ($\vee$) to express non-deterministic outcome --- as from
quantale-based interpretations of ILL \cite{abram93}. The formula $G_1
\plus G_2$ represents a graph that can match two alternative patterns
--- hence a potential situation of conflict in rule application.  The
formula $G_1 \vee G_2$ represents a graph that may have been obtained
in two different ways --- hence a situation of non-determinism.

\begin{figure}
\centering{
\includegraphics*[scale=0.65]{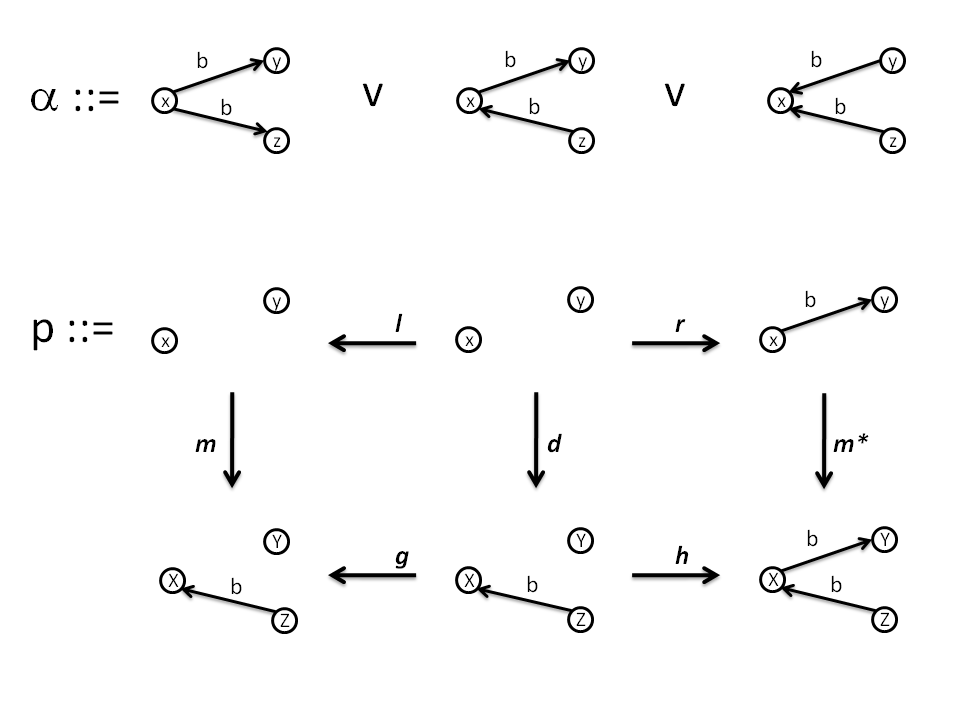}
\caption{Transformation example}\label{Transformation example}}
\end{figure}

Negative constraints can be expressed using the intuitionistic-style
negation $\ineg$. The formula $\ineg \alpha$ expresses the fact that
$\alpha$ must never be reached --- in the sense that reaching it
implies an error. In a weaker sense, the system satisfies the
constraint if $\alpha$ does not follow from the specification. To make
an example (Fig. 1), given

\[\alpha \defeq \dynex x y z:A. (b(x,y) \otimes b(x,z)) \lor
(b(x,y) \otimes b(z,x)) \lor (b(y,x) \otimes b(z,x))\]

the formula $\ineg \alpha$ says that in the system there must be no
element of type $A$ which is bound with two distinct ones (graphically
represented in the upper part of the picture). The transformation rule
in fig. 1 can be represented with $\forall x y:A.  \mathbf{1}
\multimap b(x,y)$; the initial graph with $ \dynex x y z:A.  b(z,x)$.
These two formulas specify our system. The graph transformation
determined by the application of the rule to the initial graph can be
expressed in terms of logic consequence as

\[p \defeq \dynex x y z:A. \ b(z,x), \forall x_1 x_2:A. \ \mathbf{1}
\multimap b(x_1,x_2) \Vdash \dynex x y z:A. \ b(z,x) \otimes b(x,y)\]

When the constraint $\ineg \alpha$ is added to the premises, a
contradiction follows --- as $\alpha$ already follows from the
specification. 


\section{An algebraic presentation of DPO transformation of
  hypergraphs}

Let $V$ be an infinite set of nodes $n_1, n_2, \ldots$ typed in
$\mcal{V}$, and $E$ an infinite set of edges $e_1, e_2, \ldots$ typed
in ${\cal E}$, as before.  In general, we assume typing to be implicit
--- each element $x$ associated to its type by $type(x)$.  Making type
explicit, we use $A,B,\ldots$ for node types. 




\subsection{Graph expressions}

We introduce a notion of \emph{constituent} 

$$ C = \ e(n_1,\ldots,n_k) \mid \msf{Nil} \mid C_1 \parallel C_2
\mid \upnu n. C $$


where $e(n_1,\ldots,n_k)$ is an edge component with $type(e) =
L_e(A_1,\ldots,A_k)$ when $type(n_1) = A_1, \ldots, type(n_k) = A_k$,
where $\msf{Nil}$ is the empty graph and $C_1
\parallel C_2$ is the parallel composition of components $C_1$ and
$C_2$, and where $\upnu n. C$ is obtained by restricting node name $n$
in $C$.

We say that a constituent is \emph{normal} whenever it has form $\upnu
\overline{n}. G$, where $\overline{n}$ is a (possibly empty) sequence
of node names, and $G$ is either $\msf{Nil}$ or else it does not
contain any occurrence of $\msf{Nil}$.


Given a constituent $C$, the \emph{ground components} of $C$ are the
nodes and the edge components that occur in $C$. We say that
$\mbox{\it fn}(C)$ are the \emph{free} nodes (unrestricted), $bn(C)$
are the \emph{bound} nodes (restricted), and the set of all nodes is
$n(C) \defeq \mbox{\it fn} (C) \cup bn(C)$. We denote by $cn(C)$ the
\emph{connected} nodes of $C$, i.e.  those which occur in ground
components of $C$.  We say that $ibn(C) \defeq bn(C) / cn(C)$ are the
\emph{isolated} bound nodes of $C$.  

A \emph{graph expression} is a pair $E = X \grax C$ where $X \subseteq
V$ are nodes and $C$ is a constituent such that $\mbox{\it fn}(C)
\subseteq X$. We call $X$ the \emph{interface} of $E$, or the free
nodes of $E$. The nodes of $E$ are $n(E) \defeq X \cup bn(C)$.  The
isolated free nodes are $\mbox{\it ifn}(E) \defeq X / \mbox{\it
  fn}(C)$. The isolated nodes of $E$ are $i(E) \defeq \mbox{\it
  ifn}(E) \cup ibn(C)$.  In general, $X = \mbox{\it fn}(E) \defeq
\mbox{\it ifn}(E) \cup \mbox{\it fn}(C)$, and $n(E) = i(E) \cup
cn(C)$.  We can say that graph expression $E$ is \emph{ground}
whenever $bn(C) = \emptyset$, that $E$ is \emph{weakly closed}
whenever $\mbox{\it fn}(C) = \emptyset$, that $E$ is \emph{closed}
whenever $X = \emptyset$, that $E$ is normal whenever $C$ is normal.
For simplicity, we are going to identify closed graph expressions with
their constituents.


Let $E_1 = X_1 \grax C_1, E_2 = X_2 \grax C_2$ be graph expressions in
the following.  Structural congruence between $E_1$ and $E_2$, written
$E_1 \equiv E_2$, holds iff $X_1 = X_2$ and $ C_1 \equiv C_2$, where $\equiv$
is defined over constituents according to the following axioms.
\begin{itemize}
\item The parallel operator $\parallel$ is associative and
  commutative, with \msf{Nil} as neutral element.
\item $\upnu n. \ C \equiv \upnu m. \ C [m/n]$,
  if $m$ does not occur free in $C$.\\
  $\upnu n. \upnu m. C \equiv \upnu m. \upnu n. C$ \\
  $\upnu n. (C \parallel C') \equiv C \parallel (\upnu n. C')$
  if $n$ does not occur free in $C$
\end{itemize}

We do not require $\upnu n. C \equiv C$ for $n$ not occurring free in
$C$ (we can also say that we do not require $\upnu$ to satisfy
$\eta$-equivalence). This allows us to keep isolated nodes into
account.

For $E= X \vDash C$, we denote by $ec(E)$ the edge components of $C$,
and by $gc(E) = n(E) \cup ec(E)$ the set of the ground components of
$E$. It is not difficult to see the following, with respect to $E_1$
and $E_2$.

\begin{obsA} \label{obs:1} $E_1 \equiv E_2$ if and only if $\mbox{\it
    fn}(E_1) = \mbox{\it fn}(E_2)$ and there is a renaming $\sigma$ of
  $bn(E_1)$ such that $gc(E_1)\sigma = gc(E_2)$.
\end{obsA}

One can also see that for each graph expression there is a congruent
normal one, and that congruent normal expressions are the same up to
reordering of prefix elements and ground components. 

We say that $E_1$ is a \emph{heating} of $E_2$ (conversely, that $E_2$
is a \emph{cooling} of $E_1$), and we write $E_1 << E_2$, whenever
there is a graph expression $E_3 = X_3 \vDash \upnu \overline{n}. C_1$
such that $X_3 = X_2/\{\overline{n}\}$ (for $\overline{n}$ possibly
empty sequence of node names) and $E_3 \equiv E_2$ --- i.e. when $E_2$
can be obtained from $E_1$ modulo congruence by restricting node
names. Therefore, intuitively, $E_1$ is one of the smallest patterns
$E_2$ can match with, and conversely, $E_2$ is one of the largest
graphs that can match with $E_1$. This essentially means that,
although not congruent, as operationally different, $E_1$ and $E_2$
share the same structure.



An abstract hypergraph in the sense of section 2.1 is represented by
an equivalence class of graph expressions up to structural congruence.
Intuitively, the free names correspond to nodes in the interface while
bound names represent internal nodes.

We will often refer to these equivalence classes as graphs, while
reserving the term hypergraph for the real thing. We say that a graph
expression represents a graph (is a \emph{representative} of the
graph) when it belongs to the equivalence class. A graph is (weakly)
closed whenever it is represented by a (weakly) closed graph
expression. Clearly, every closed graph has a closed normal
representative.

It is not difficult to see that a graph can also be represented as the
class of all the heatings of its representatives --- leading to a
semantics based on partial orders rather than equivalence relations.
We will refer to heatings of graph expressions that represent
subgraphs of a given expression $E$ as \emph{heating fragments} of $E$
(conversely, their \emph{cooling compound}).


\subsection{Transformation rules}

In order to represent transformation rules we need to deal with the
matching of free nodes. To this purpose we introduce variables $x,y,
\ldots$ ranging over nodes, substitution of nodes for free variables
($E[m/x]$, where $m$ does not contain occurrences that become bound),
variable binding (by $\Lambda$) and application.




For $E_1,E_2$ closed graph expression, $E_1 \Rrel{} E_2$ denotes the
transformation that goes from $E_1$ to $E_2$.  Given graph expressions
$E_1 = \ K \grax L$ and $E_2 = \ K \grax R$ sharing the same interface
and no free isolated nodes, we represent the transformation rule $\pi
(p) = \ L \lrel l K \rrel r R$ by the \emph{rule expression} $\Lambda
\overline {x}.  L \Rrel{p} R$, where $\overline{x} = x_1,\ldots,x_k$
is a sequence of variables associated to the node names in $K$.
Essentially, we represent rules by replacing each free node with a
bound variable.

Given a closed graph representative $G$, a match for $\pi(p)$ in $G$
(as pictured in section 2.1) is determined by a graph homomorphism $d:
K \to n(G)$ which determines the left hand-side morphisms $m: L \to
G$, with components $m_v: bn(L) \to n(G)$ and $m_e: ec(L) \to ec(G)$,
as well as right hand-side morphism $m^*: R \to H$, with components
$m^{*}_v: bn(R) \to n(H)$ and $m^{*}_e: ec(R) \to ec(H)$.





The dangling edge condition means that if $n$ is in the domain of
$m_v$ and occurs in component $c$, then $c$ must be in the domain of
$m_e$.  The identification condition requires that $m_v$ and $m_e$ are
injective, and that the images of $d$ and $m_v$ are disjoint. The
injectivity of $m^{*}_v$ and $m^{*}_e$ follows, as well as the
disjointness of the images of $d$ and $m^{*}_v$.

The injective components can be represented in terms of inclusion,
whereas the interface morphism $d$ can be represented in terms of
substitution, i.e. we represent $d$ by $[\overline{n} \lrel d
\overline{x}] = [x_1/n_1, \ldots x_k/n_k]$, where $\overline{n} =
\{n_1,\ldots,n_k\} \subseteq n(G)$.  The following operational rule
(application schema) represents the application of the transformation
rule $p$ with match $m$ (determined by $d$)

$$ \infer [\Rrel{\langle p,m \rangle}] { G \Rrel{p,d} H }
{ \begin{array}{llll} \pi(p) = \Lambda \overline{x}. L \Rrel{p} R & G
    \equiv \upnu \overline{n}. L [\overline{n} \lrel d \overline{x}]
    \parallel C &
    H \equiv \upnu \overline{n}. R [\overline{n} \lrel d
    \overline{x}]
    \parallel C
  \end{array} }
$$

where $G$ is a closed graph expression --- and therefore $H$ is, too.

\begin{obsA} \label{obs:3} The application schema satisfies the DPO
  conditions.

  Let $L' = L[\overline{n} \lrel d \overline{x}]$, $R' =
  R[\overline{n} \lrel d \overline{x}]$. The definition and the
  injectivity of component morphisms $m_v,m_e,m^{*}_v,m^{*}_e$ follows
  from the inclusion of $L'$ and $R'$ as subexpressions in
  refactorings of $G$ and $H$, respectively.  The disjointness
  condition holds by the fact that the variables in $\overline{x}$ are
  substituted with nodes that are free in $L'$ and $R'$, and therefore
  cannot be identified with bound nodes in either constituent. The
  dangling edge condition holds by the fact that, for each node $n \in
  bn(L')$, edge components depending on $n$ can only be in $ec(L')$.
\end{obsA}



\section{Linear lambda-calculus}
 
We rely on a constructive presentation of intuitionistic linear logic,
based on the labelling of logic formulas, in a way that gives rise to
a form of $\lambda$-calculus. Linear $\lambda$-calculus
\cite{abram93,BeEA93,CePf02,Pf02} has been introduced in association
with intuitionistic linear logic and with the notion of linear
functions, by interpreting linearity as consumption of arguments.
Linear implication ($\multimap$) can be used to type linear functions,
as much as intuitionistic implication ($\to$) is used to type generic
ones.

We rely on a two-entry sequent presentation of linear logic
\cite{Pf94,Pf02}, and we follow the convention to use different sorts
of variable identifiers for linear resources ($u,v,\ldots$) and
non-linear ones ($p,q,\ldots$).  We denote linear abstraction by
$\lnlambda$ (with $\caret$ for linear application), to distinguish it
from standard one ($\lambda$) --- though the difference between the
two can actually be determined by whether the abstraction variable is
linear.  For the purpose of the translation, we find it further useful
to distinguish individual variables ($x,y,\ldots$, non-linear), and
node variables ($m,n,\ldots$, linear).  Whether $\lambda$ is typed by
$\forall$ depends on whether the abstraction is over an individual
variable that occurs in the type. We use \emph{let} expressions to
abstract over patterns. We assume standard forms of $\alpha$-renaming,
$\beta$- and $\eta$-congruence for $\lambda$ and $\lnlambda$ (with
linearity check for the latter).

$N::\alpha$ is a typing expression (typed term) where $N$ is a term
(the label) and $\alpha$ is a logic formula (the type). Two-entry
sequents have form $\Gamma; \Delta \vdash N::\alpha$, where $\Delta$
is a multiset of typed linear variables (linear context), with
$\Delta_N \subseteq \Delta$ a multiset of typed node variables, and
$\Gamma$ is a multisets of typed non-linear variables (non-linear
context), with $\Gamma_I \subseteq \Gamma$ a multiset of typed
individual variables. We use sequence notation --- modulo permutation
and associativity, and a dot ($\cdot$) for the empty multiset.

A natural deduction systems is given by a set of axioms and a set of
primitive inference rules, each associated as either introduction or
elimination operational rule to a logical operator. A sequent is
provable, and represents a typing derivation, when it can be derived
from the axioms by means of inference rules.  We say that a rule

$$\infer[]{\Sigma}{\Sigma_1 \ldots \Sigma_n}$$

is derivable whenever it can be proved that, if
$\Sigma_1,\ldots,\Sigma_n$ are provable sequents, then also $\Sigma$
is. When we ``forget'' all about labels we are left with logic
formulas and the consequence relation --- then we use $\Vdash$ instead
of $\vdash$.


\subsection{A system with restriction}

We consider a system with standard propositional intuitionistic linear
operators $\multimap, \otimes, \One, \top, \bot, \to, \lor, \land, !$
and standard universal quantifier $\forall$.  Each of these can be
associated to a linear $\lambda$-calculus operator
\cite{abram93,Pf02}. We also allow for syntactical type equality (=),
stronger than linear equivalence ($\lequiv$, which can be defined in
terms of $\multimap$ and $\land$). We assume standard rules for $=$.
However, we only need to prove instances of type equality arising from
substitution as side conditions, and we do not actually use the
proof-term --- therefore for simplicity we associate $=$ to an axiom
and a dummy term $\msf{nil}_{=}$.
 
We extend this system by adding resource-bound existential
quantification ($\dynex$) and an auxiliary modifier to express
reference ($\downmapsto$). The extension is meant to answer two
issues.  First --- nodes need to be treated linearly from the point of
view of transformation, though their names occur non-linearly in graph
expressions. Second --- we need to associate a type to name
restriction in the context of graph expressions. The
resource-boundedness of $\dynex$ makes it possible to treat nodes
linearly, whereas the freshness conditions on $\dynex$ and
$\downmapsto$ make it possible to interpret operationally $\dynex$ as
restriction type. 

The modifier $\downmapsto$ is meant to express reference of an
individual variable (a node name) to a linear one (a node) as part of
the node type. The typed linear variable $n::\alpha \downmapsto x$ is
referred to by the typed non-linear one $x::\alpha$ --- we will also
say that $n$ is a \emph{reference} variable, and that $x$ is the
\emph{referring} variable in $\alpha \downmapsto x$.  We require, as
operational constraint, that each individual variable may occur as
referring variable no more than once in the linear context of a
sequent (\emph{uniqueness constraint}). This constraint entails that
the reference relation between reference variables and individual free
variables is one-to-one, and also that reference variables can only be
linear.

We use $\lnvareps$ to denote the restriction-like operator associated
with $\dynex$, that can be defined as

$$ \lnvareps (n|y). M :: \dynex x: \alpha. \beta \ \defeq 
\ y \otimes n \otimes M $$

where $n::\alpha \downmapsto y$ and therefore $y$ refers to $n$. The
definition of $\lnvareps$ is essentially based on that of
proof-and-witness pair associated with the interpretation of
existential quantifier, in standard $\lambda$-calculus
\cite{ProofTh00} as well as in its linear version \cite{CePf02,Pf02}.

The inference rules guarantee that there is a one-to-one relation
between referring variables in the context of a sequent and variables
that may be bound by $\dynex$ (\emph{naming property}), under the
assumption that $\dynex$ does not occur in the axioms. The property is
preserved by the $\dynex$ elimination rule --- similar to the standard
existential quantifier rule, requiring that the instantiated term (a
referring variable) as well as the associated reference are fresh
variables.  We force the naming property to be preserved by the
$\dynex$ introduction rule, by requiring that the new bound variable
replaces all the occurrences of the instantiated term in the
consequence of the derivation (\emph{freshness condition} of $\dynex$
introduction). The naming property follows for any provable sequent,
under the given assumption, by induction over proofs. Given the
uniqueness constraint, it also follows that there is a one-to-one
relation between reference variables in the context of a sequent and
variables that may be bound by $\dynex$ (\emph{linear naming
  property}).

The freshness condition of $\dynex I$ is expressed formally, in terms
of substitution and syntactical type equality ($\Gamma, x::\beta;
\cdot \vdash \msf{nil}_{=}:: \alpha \# (x,y)$).  
In fact, as we define $\alpha \# (x,y) \ \defeq \ (\alpha[y/x])[x/y] =
\alpha$, the typed term $\msf{nil}_{=} \alpha \# (x,y)$ can be used to
express that $y$ does not occur free in $\dynex x.\alpha$. However, this is
essentially just the formalisation of a side condition for the rule.
From the freshness condition it also follows that the formula $\dynex
x.  \alpha$ obtained by $\dynex$ introduction is determined, modulo
renaming of bound variables, by the instance $\alpha[y/x]$ in the
hypothesis.

The linear naming property ensures that $\dynex$ can be used to bind
free variables, hiding them, though without allowing any derivation of
instantiations that can alter irreversibly the structure of the
formula, and that therefore these variables can be treated as names,
preserved through inference --- and moreover, that these names are
associated with linear resources. In chemical terms, with reference to
section 3.1, borrowing a suggestion from \cite{Berry90}, $\dynex$
allows us to understand derivation as cooling process.

A normal proof is intuitively speaking one in which there are no
detours --- no operators that are introduced to be thereafter
eliminated. A system is normalising whenever every provable sequent
has a normal proof. All provable sequents in ILL have normal proofs
\cite{abram93} and this result can be extended to the logic with
standard quantifier (see \cite{Pf02}, though an unpublished). A proof
that our system is complete with respect to normal proofs goes beyond
the scope of the present paper.  However, it is informally arguable
that completeness holds essentially, by translation to a sequent
calculus system, for which it is comparatively easier to see that the
fragment $\otimes,\multimap,\One,\forall,\dynex$ enjoys the cut
elimination property, closely associated with proof normalisation.

\subsection{Quantification and DPO properties}

We have introduced resource-bound quantification in order to express
more easily the injective character of the pattern-matching morphism
components associated with deletion and creation of graph elements. It
is not difficult to see that the following, closely associated
properties hold --- in clear contrast with what happens with standard
existential quantification.

\begin{obsA} \label{obs:4}
(1) \ $\nVdash (\dynex x:\beta. \
  \alpha(x,x)) \ \multimap \ \dynex x y:\beta. \ \alpha(x,y)$

the resource associated to $x$ cannot suffice for $x$ and $y$.

(2) \ $\nVdash \forall x:\beta. \ \beta \downmapsto x \otimes 
\alpha(x,x) \ \multimap \ \dynex y:\beta.
\alpha(y,x)$

$y$ and $x$ should be instantiated with the same term --- but this is
prevented by the freshness condition in $\dynex$ introduction

(3) \ $\nVdash (\dynex y x:\beta. \ \alpha_1(x) \otimes \alpha_2(x)) \
\multimap \ (\dynex x:\beta. \alpha_1(x)) \otimes \dynex x:\beta.
\alpha_2(x)$

the two bound variables in the consequence require
distinct resources and refer to distinct occurrences
\end{obsA}

In particular, (1) and (2) can be regarded as a properties associated
with the identification condition, whereas (3) has a more general
structure-preserving character.

The following properties show a relationship between linear
equivalence and the congruence relation defined in section 3.1.

\begin{obsA} \label{obs:5} $\dynex$ satisfies properties of
  $\alpha$-renaming, exchange
  and distribution over $\otimes$, i.e. \\

$\Vdash (\dynex x:\alpha. \beta(x)) \ \lequiv \ (\dynex y:\alpha.
\beta(y))$

$\Vdash (\dynex x y:\alpha. \gamma) \ \lequiv \ (\dynex y x. \gamma)$

$\Vdash (\dynex x:\alpha. \beta \otimes \gamma(x)) \ \lequiv \ (\beta
\otimes \dynex x:\alpha. \gamma(x)) \qquad$ ($x$ not in $\alpha$)
\end{obsA}

In general $\dynex$ does not satisfy logical $\eta$-equivalence, i.e.
it cannot be proved that $\alpha$ is equivalent to $\dynex x. \
\alpha$ when $x$ does not occur free in $\alpha$ (neither sense of
linear implication holds). This is useful though, in order to
represent graphs with isolated nodes. Note that, in order to match the
notion of congruence introduced for graph expressions at the term
level, term congruence in the $lambda$-calculus should be extended
with $\alpha$-renaming, exchange, and distribution over $\otimes$ for
$\lnvareps$.  However this is not needed here, insofar as we can
reason about congruence at the type level, in terms of linear
equivalence.

\subsection{Proof systems (QILL)}

\noindent $ \alpha = A \mid L(N_1,\ldots,N_n) \mid \One \mid \alpha_1
\otimes \alpha_2 \mid \alpha_1 \multimap \alpha_2 \mid ! \alpha_1 \mid
\top \mid \bot \mid \alpha_1 \plus \alpha_2 \mid \alpha \to \beta \mid
\alpha \lor \beta \mid \forall x:\beta. \alpha \mid \dynex x:\beta.
\alpha \mid \alpha \downmapsto x \mid \alpha = \alpha $\\

\noindent $ M = x \mid p \mid n \mid u \mid \msf{nil} \mid N_1 \otimes
N_2 \mid \lnvareps (N_1 | N_2). N_3 \mid \lambda x. N \mid \lambda p.
N \mid \lnlambda u. N \mid N_1 \caret N_2 \mid N_1 N_2 \mid
\msf{error}^{\alpha} \ M \mid \langle \rangle \mid \langle N_1,N_2
\rangle \mid \msf{fst} \ N \mid \msf{snd} \ N \mid \msf{case} \ N \
\msf{of} \ P_1. N_1; P_2. N_2 \mid \msf{inr}^{\alpha} \ N \mid
\msf{inl}^{\alpha} \ N \mid \msf{nil}_{=} $\\


$\msf{let} \ P = N_1 \ \msf{in} \ N_2 \ \defeq \ (\lambda P. N_2) N_1
\qquad $ where
$P$ is a variable pattern\\

$\alpha \lequiv \beta \ \defeq \ (\alpha \multimap \beta) \plus (\beta
\multimap \alpha)$ $\qquad \neg \alpha \ \defeq \ \alpha \multimap
\bot$ $\qquad \alpha \# (x,y) \ \defeq \ (\alpha [y/x])[x/y] = \alpha$



\[
\begin{array}{ll}
  \infer [Id] {\Gamma; u::\alpha \vdash u::\alpha} {}   \qquad & \qquad
  \infer [UId] {\Gamma, p::\alpha; \cdot \vdash p::\alpha} {} \\
  \\
  \infer [NId] {\Gamma, x::\alpha; n::\alpha \downmapsto x 
                \vdash n::\alpha \downmapsto x} {}    \qquad & \qquad
  \infer [Eq] {\Gamma; \cdot \vdash \msf{id}_{\alpha}::
    \alpha = \alpha} {}
\end{array}
\]

\[
\begin{array}{ll}
  \infer [\otimes I] { \Gamma; \Delta_1, \Delta_2 \vdash M \otimes N::
    \alpha \otimes \beta } { \Gamma; \Delta_1 \vdash M::\alpha & \Gamma;
    \Delta_2 \vdash N::\beta }  &
  \infer [\otimes E] { \Gamma; \Delta_1, \Delta_2 \vdash \msf{let} \ u \
    \otimes v = M \ \msf{in} \ N:: \gamma } { \Gamma; \Delta_1 \vdash M::
    \alpha \otimes \beta & \Gamma; \Delta_2, u::\alpha, v::\beta \vdash
    N::\gamma }  
\end{array}
\]

\[
\begin{array} {ll}
  \infer [\multimap I] { \Gamma; \Delta \vdash \lnlambda u:\alpha. \ M::
    \alpha \multimap \beta } { \Gamma; \Delta, u:: \alpha \vdash M:: \beta
  }  &
  \infer [\multimap E] { \Gamma; \Delta_1, \Delta_2 \vdash M \caret N::
    \beta } { \Gamma; \Delta_1 \vdash M:: \alpha \multimap \beta &
    \Gamma; \Delta_2 \vdash N:: \alpha }
\end{array}
\]

\[
\begin{array} {ll}
  \infer [\mathbf{1}I] { \Gamma; \cdot \vdash \msf{nil} :: \mathbf{1} } {}
  &
  \infer [\mathbf{1}E] { \Gamma; \Delta, \Delta' \vdash \msf{let} \ \msf{nil}
    = M \ \msf{in} \ N:: \alpha } { \Gamma; \Delta \vdash M :: \mathbf{1}
    & \Gamma; \Delta' \vdash N:: \alpha }  
\end{array}
\]

\[
\begin{array} {ll}
    \infer [\plus I] { \Gamma; \Delta \vdash \langle M, N \rangle:: \alpha
    \plus \beta } { \Gamma; \Delta \vdash M:: \alpha & \Gamma; \Delta
    \vdash N:: \beta }
  &
  \infer [\vee E] { \Gamma; \Delta, \Delta' 
    \vdash \msf{case} \ M \ \msf{of} \ \msf{inl} \ u. \ N_1; 
    \ \msf{inr} \ v. \ N_2 \ :: \gamma }  
  { \Gamma; \Delta \vdash M :: \alpha \vee \beta & 
    \Gamma; \Delta', u :: \alpha \vdash N_1 :: \gamma & 
    \Delta', v :: \beta \vdash N_2 :: \gamma } 
\end{array}
\]

\[
\begin{array} {ll}
  \infer [\plus E1] { \Gamma; \Delta \vdash \msf{fst} \ M:: \alpha } {
    \Gamma; \Delta \vdash M:: \alpha \plus \beta }  &
  \infer [\plus E2] { \Gamma; \Delta \vdash \msf{snd} \ M:: \beta } {
    \Gamma; \Delta \vdash M:: \alpha \plus \beta }  
\end{array}
\]

\[
\begin{array} {ll}
  \infer [\vee I1] {\Gamma; \Delta \vdash M:: \alpha}
  {\Gamma; \Delta \vdash \msf{inl}^\beta M :: \alpha \vee \beta}
  &
  \infer [\vee I2] {\Gamma; \Delta \vdash M:: \beta}
  {\Gamma; \Delta \vdash \msf{inr}^\alpha M :: \alpha \vee \beta} \\
\\
  \infer [\top I] { \Gamma; \Delta \vdash \langle \rangle :: \top } { } &
  \infer [\bot E] { \Gamma; \Delta, \Delta' \vdash
    \msf{error}^{\alpha} \ M:: \alpha } { \Gamma; \Delta \vdash M:: \bot} 

\end{array}
\]

\[
\begin{array} {ll}
\infer [! I] { \Gamma; \cdot \vdash \ !M :: \ !\alpha } { \Gamma; \cdot
  \vdash M:: \alpha } &
\infer [! E] { \Gamma; \Delta_1, \Delta_2 \vdash \msf{let} \ p = M
  \  \msf{in} \ N:: \beta } { \Gamma; \Delta_1 \vdash M:: \ !\alpha &
  \Gamma, p:: \alpha; \Delta_2 \vdash N:: \beta }
\end{array}
\]

\[
\begin{array} {ll}
  \infer [\to I] { \Gamma; \Delta \vdash \lambda p. \ M:: \alpha
    \to \beta } { \Gamma, p::\alpha; \Delta \vdash M:: \beta } &
  \infer [\to E] { \Gamma; \Delta \vdash M N :: \beta } { \Gamma; \Delta
    \vdash M:: \alpha \to \beta & \Gamma; \cdot \vdash N:: \alpha }  
\end{array}
\]

\[
\begin{array} {ll}
\infer [\forall I] { \Gamma; \Delta \vdash \lambda x. \ M:: \forall
  x:\beta. \ \alpha } { \Gamma, x::\beta ; \Delta \vdash M:: \alpha }  &
\infer [\forall E] { \Gamma; \Delta \vdash M N:: \alpha [N / x] } {
  \Gamma; \Delta \vdash M:: \forall x:\beta. \ \alpha & \Gamma; \cdot
  \vdash N:: \beta }
\end{array}
\]

\subsubsection*{Resource-bound quantifier}

\[
\begin{array} {l}
  \infer [\dynex I]
  { \Gamma; \Delta, \Delta' \vdash \lnvareps (n | y).
    M:: \dynex x: \beta. \alpha } { \Gamma; \Delta \vdash M::
    \alpha [y / x] & \Gamma; \Delta' \vdash
    n::\beta  \downmapsto y &
    \Gamma, x::\beta; \cdot \vdash \msf{nil}_{=}:: \alpha \# (x,y) }  \\
  \\
  \infer [\dynex E] { \Gamma; \Delta_1, \Delta_2 \vdash \msf{let} \ \lnvareps
    (n|x). v = M \ \msf{in} \ N:: \gamma }
  { \Gamma; \Delta_1 \vdash M:: \dynex x:
    \beta. \ \alpha & \Gamma, x::\beta; \Delta_2,
    n::\beta \downmapsto x, v::\alpha \vdash N:: \gamma }  
\end{array}
\]

\section{Linear encoding of GTS}

We are going to define a translation of graph expressions to typing
derivations.  Intuitively, the translation is based on a quite
straightforward mapping of graph expressions into proof terms, with
$\msf{Nil}$ mapped to $\msf{nil}$, $\parallel$ to $\otimes$, and
$\upnu$ to $\lnvareps$.  However, we need to distinguish nodes as
ground components (nodes) from node occurrences in constituents (node
names).  Given $E = X \vDash C$, we can translate a node $n \in X$
with $type(n) = A$ as $n::A\downmapsto x$ (\emph{typed node}), and the
occurrences of $n$ in $C$ as $x_n::A$, where $A$ is an unbounded
resource type (therefore equivalent to $!A$).

Semantically, it is more convenient to take edge components as
primitive, rather than edges. In principle, we can introduce a notion
of \emph{edge interface} as linear resource, $e:: \forall
x_1:A_1,\ldots, x_k:A_k.  L_e(x_1,\ldots x_k)$, translate an edge type
$L_e(A_1,\ldots,A_k)$ as $\forall x_1:A_1,\ldots, x_k:A_k.
L_e(x_1,\ldots x_k)$, and a component $e(n_1,\ldots, n_k)$ as $c_e = e
\ x_1 \ \ldots x_k$. For all its functional clarity, however, the
notion of edge interface is hard to place in GTS.  Therefore, we
prefer to introduce the notion $c_e::L(x_1,\ldots x_n)$ of \emph{typed
  edge component} as primitive, which can be translation of the
original component under the premises $x_1::A_1, \ldots x_k::A_k$.
Following this approach, component connectivity does not result from
the term, rather from the type.


We call \emph{graph formulas} those in the $\mathbf{1},\otimes,\dynex,
\downmapsto$ fragment of the logic containing only primitive graph
types (node and edge types). We say that a graph formula $\gamma$ is
in normal form whenever $\gamma \ = \ \dynex (\overline{x:A}). \
\alpha$, where either $\alpha = \mathbf{1}$ or $\alpha =
L_1(\overline{x}_1) \otimes \ldots \otimes L_k(\overline{x}_k) $, with
$\overline{x::A}$ a sequence of typed variables. The formula is closed
if $\overline{x}_{i} \subseteq \overline{x}$ for each $1 \leq i \leq
k$. A \emph{graph context} is a multiset of typed nodes and typed edge
components.

A \emph{graph derivation} is a valid sequent $\Gamma; \Delta \vdash
N:: \gamma$, where $\gamma$ is a graph formula, $\Delta$ is a graph
context and $\Gamma$ contains only individual variables. A graph
derivation uses only axioms and the introduction rules $\mathbf{1} I$,
$\otimes I$, $\dynex I$ --- therefore it is trivially normal.

We can now define formally the translation as function $\llbracket
\rrbracket$ from graph expressions to typing derivations.
We use the notation $AxiomName \ [\Gamma;; \ form]$ to abbreviate
axiom instances and deduction rules with empty hypothesis (by giving
the non-linear context and the principal formula, if there is one),
and $RuleName \ [ hyp_1;; \ \ldots;; \ hyp_n ]$ to abbreviate instances
of inference rules (by giving the hypothesis). We also define
$MainType( \Gamma;\Delta \vdash N::\alpha ) = \alpha$, $MainTerm(
\Gamma;\Delta \vdash N::\alpha ) = N$, and $LinearContext(\Gamma;
\Delta \vdash N::\alpha) = \Delta$ as auxiliary functions.






\subsubsection*{Constituents}

$$
\begin{array}{l}
  \llbracket e_i (m, \ldots, n):L_i(A_m,\ldots,A_n) \rrbracket \ \defeq \ Id \ [\Gamma;; \quad c_i  :: L_i(x_m,\ldots,x_n)]  \\
  \llbracket \msf{Nil} \rrbracket \ \defeq \ \mathbf{1}I \ [\Gamma]  \\
  \llbracket M \parallel N \rrbracket \ \defeq \ \otimes I \ [\llbracket M \rrbracket;; \quad \llbracket N \rrbracket]   \\
  \llbracket \upnu n:A. N \rrbracket \ \defeq \ \dynex I \ [ \ \llbracket N \rrbracket;; \\ 
  \qquad \qquad \qquad \quad \quad NId \ [\Gamma;; \ n::A \downmapsto x_n];; \\ \qquad \qquad \qquad \quad \quad \Gamma, y::A;\cdot \vdash \msf{nil}_{=}:: MainType(\llbracket N \rrbracket)[y/x_n] \# (y,x_n) \ ]
\end{array}
$$

\subsubsection*{Graph interfaces}

$$
\begin{array}{l}
  \llbracket n:A \rrbracket \ \defeq \ NId \ [\Gamma;; \ n :: A \downmapsto x_n]  \\
  \llbracket \{ n:A \} \rrbracket \ \defeq \ \llbracket n:A \rrbracket \\
  \llbracket \{ n_1:A_1 \} \cup X \rrbracket \ \defeq \ \otimes I \ [\llbracket \{ n_1:A_1 \} \rrbracket;; \ \llbracket X \rrbracket]
\end{array}
$$

\subsubsection*{Graph expressions}

$$
\begin{array}{l}
  \llbracket X \vDash C \rrbracket \ \defeq \ \otimes I \ [\llbracket X \rrbracket_I;; \ \llbracket C \rrbracket]
 \end{array}
$$

\subsection{Properties of the translation}


We first consider the following induced mapping, taking graph
expressions into QILL formulas ($\llbracket \rrbracket^{T}$), and into
multisets of typed variables associated to ground components
($\llbracket \rrbracket^{C}$). In fact, let $\llbracket E
\rrbracket^{T} = MainType \llbracket E \rrbracket$ and $\llbracket E
\rrbracket^{C} = LinearContext \llbracket E \rrbracket$.

\begin{obsA} \label{obs:10} 1) $\llbracket \rrbracket^T$ results in an
  extension of the original typing of nodes and edges, based on the
  association of $\otimes$ with $\parallel$, $\mathbf{1}$ with
  $\msf{Nil}$, and $\dynex$ with $\upnu$, where the
  free connected nodes are represented as free variables occurring in the consequence (which, by definition of $\llbracket \rrbracket$, are all referring), whereas other free referring variables represent free isolated nodes.  \\
  2) $\llbracket E \rrbracket^{C} = \Delta$ determines a bijection
  between $\Delta$ and $gc(E)$ --- dependant types contain the
  information about basic graph types and component dependencies,
  whereas terms preserve component identity.
\end{obsA}

\begin{propA} \label{prop:1} There is an isomorphism between graph
  expressions and graph derivations.

  For each $E$ graph expression, $\llbracket E \rrbracket = \Gamma;
  \Delta \vdash N:\gamma$ defines a graph derivation. By construction,
  $N$ and $\Gamma$ are as required, $\llbracket E \rrbracket^{T}$
  gives a graph formula, $\llbracket E \rrbracket^{C}$ a graph
  context.  Vice-versa, for each graph derivation $\delta = \Gamma;
  \Delta \vdash N:\gamma$, one can define a graph expression $E$ such
  that $\llbracket E \rrbracket = \delta$, relying on Obs.
  \ref{obs:10}.
\end{propA}

\begin{propA} \label{prop:2} There is an isomorphism between graphs
  and graph formulas modulo linear equivalence.

  Given graph expressions $M,N$, if $M \equiv N$ then $\Vdash
  \llbracket M \rrbracket^{T} \lequiv \llbracket N \rrbracket^{T}$.
  This follows from the monoidal characterisation of $\otimes$ and
  Obs. \ref{obs:5}.

  On the other hand, assume $\gamma_1,\gamma_2$ are graph formulas and
  $\Vdash \gamma_1 \lequiv \gamma_2$. Then, for each graph expressions
  $E_1,E_2$ such that $\gamma_1 = \llbracket E_1 \rrbracket^{T}$,
  $\gamma_2 = \llbracket E_2 \rrbracket^{T}$, it holds $E_1 \equiv
  E_2$. By property of linear equivalence, there is a graph derivation
  $\delta_1 = \Gamma; \Delta \vdash N_1::\gamma_1$ iff there is a
  graph derivation $\delta_2 = \Gamma; \Delta \vdash N_2::\gamma_2$.
  By Prop. \ref{prop:1}, there are graph expressions $E_1,E_2$ such
  that $\llbracket E_1 \rrbracket = \delta_1, \llbracket E_2
  \rrbracket = \delta_2$. From Obs. \ref{obs:10}(2), $gc(E_1) =
  gc(E_2)$. Since $\gamma_1$ and $\gamma_2$ are equivalent they share
  the same free variables, and so do $E_1$ and $E_2$, by Obs.
  \ref{obs:10}(1). Hence follows $E_1 \equiv E_2$, by Obs.
  \ref{obs:1}.
\end{propA}

The propositions above state that there is a Curry-Howard isomorphism
between graph expressions and graph derivations on one side, and
between graphs and QILL formulas modulo equivalence on the other. They
also state that our translation of graph expressions is adequate with
respect to their congruence.

By an argument similar to that of Prop. \ref{prop:2} and the
definition of heating (section 3.1), we can prove also the following.

\begin{obsA} \label{prop:20} Given graph expressions $M,N$, the
  sequent $\Vdash \llbracket M \rrbracket^T \multimap \llbracket N
  \rrbracket^T$ is provable if and only if $M$ is a heating of $N$.
\end{obsA}

This observation has wider semantical consequences, by noting that all
the inference rules involved in graph derivation, if read backward,
lead to graph derivations that represent heating fragments of the
graph expression represented by the conclusion.

\subsection{DPO transformations}

We can now shift from congruence of graph expressions to reachability
in a GTS, extending the translation to deal with graph transformation.
We consider transformation up to isomorphism, and therefore we start
from the type level, relying on Prop.  \ref{prop:2} --- i.e. we define
directly the map $\llbracket \rrbracket^{T}$ from graph expressions to
QILL formulas. We do this by associating transformation to linear
implication, and the binding of node variables in rule interfaces to
universal quantification.

\[
\begin{array}{l}
  \llbracket M \Rrel{} N \rrbracket^{T} \defeq \llbracket M \rrbracket^{T} \multimap \llbracket N \rrbracket^{T} \\
  \llbracket \Lambda x:A. N \rrbracket^{T} \defeq \forall x: A. \llbracket N \rrbracket^{T} \\
\end{array}
\]

Transformation rules are meant to be primitive in a GTS, so they can
be introduced as premises (as with nodes and edge components). They
have to be regarded as unbounded resources, in order to account for
their potentially unlimited applicability, and moreover they must be
associated with closed formulas (as there are neither free nodes nor
free variables in transformation rule expressions). Reasoning at an
abstract level, it seems appropriate to forget about proof terms and
consider only the types of the formulas associated with the graph
expressions in the algebraic definition of the rule.

The translation of a rule $\pi(p) = \Lambda \overline{x}. L \Rrel{} R$
can therefore be defined as follows

\[
\begin{array}{l}
  \llbracket \pi(p) \rrbracket \ \defeq \ FId \ [\Gamma;; \quad p::\forall \overline{x:A_x}. \llbracket L \rrbracket^{T} \multimap \llbracket R \rrbracket^{T} ]
\end{array}
\]



At an intuitive level, in terms of natural deduction and of a proof
built from the bottom, the application of rule $p$ to a graph $G$
involves deriving the matching subgraph $L'$ from $gc(L') \subseteq
gc(E)$. The application of $p$ to $L'$ can be understood as an
instantiation of the rule interface, corresponding to $\forall$
elimination proof steps; followed by an application of the
instantiated rule to $L'$, corresponding to a $\multimap$ elimination
step, and resulting into a conclusion that represents $R'$; followed
by a graph derivation of $H$ from premises that represent $gc(R') \cup
(gc(E) / gc(L'))$.

From a more goal-oriented perspective, assuming normalisation, the
application of $p$ to $G$ can be seen as a process leading to a
heating fragment of $G$, which in turn is a heating of rule match.

More formally, the application of $p$ to a closed graph formula
$\alpha_G = \dynex \overline{y:A_y}. \beta_G$ determined by morphism
$m$ relies on the fact that the following application schema is a
derivable rule (proof along the lines of the above intuitive
explanations)

$$
\infer [\Rrel{p,m}] { \Gamma;  \forall \overline{x:A_x}.
    \alpha_L \ \multimap \ \alpha_R \Vdash \alpha_G \multimap \alpha_H }
{
  \begin{array}{llll} 
    \Gamma; \cdot \Vdash \alpha_G \lequiv \alpha_{G'} & \alpha_{G'} =
    \dynex \overline{z:A_z}. \alpha_L [\overline{z:A_z} \lrel d
    \overline{x:A_x}]
    \otimes \alpha_C \\
    \Gamma; \cdot \Vdash \alpha_H \lequiv \alpha_{H'} & \alpha_{H'} =
    \dynex \overline{z:A_z}. \alpha_R
    [\overline{z:A_z} \lrel d \overline{x:A_x}]
    \otimes \alpha_C
  \end{array}
}
$$


where the interface morphism $d$ associated with $m$ is represented by
the multiple substitution $[\overline{z:A_z} \lrel d
\overline{x:A_x}]$, with $\overline{z:A_z} \subseteq
\overline{y:A_y}$.

Along these lines, it is possible to see that a hypergraph
transformation system $\mcal{G} = \langle TG,P,\pi,G_0 \rangle$ can be
translated to QILL, and that it is possible to obtain an adequacy
result for QILL with respect to reachability in GTS

\begin{propA} \label{obs:20} The translation is complete and correct
  with respect to reachability in DPO-GTS (restricting to rules with
  only nodes in the interface).

  For the completeness side --- given that we can represent every
  graph, it is not difficult to see that we can also simulate every
  rule application in QILL.

  For the correctness side, we need to show that every provable
  sequent expressing a transformation from a graph formula to another
  one by means of transformation rule formulas, can be simulated in
  the algebraic formalism. We can focus on a single transformation
  rule application as inductive step case, i.e. considering a sequent
  $\Gamma; R, G_1 \Vdash G_2$ where $G_1,G_2$ are graph formulas and
  $R$ a transformation rule formula. Assuming that we have a normal
  proof, we can argue that each backward step gives heating fragments
  of $G_2$ (introduction rules) and of $G_1$ (elimination rules) ---
  therefore preserving structure. It is a matter of routine ---
  induction on number of variables and graph nodes --- to show that
  the instantiations of $R$ correspond, up to isomorphism, to the
  matches of the corresponding algebraic tranformation rule $R'$.
  Therefore the sequent is provable only if the algebraic graph $G_2'$
  can be reached from the algebraic graph $G_1'$ by application of
  $R'$.
\end{propA}

The following may give an idea of the level of expressiveness.

\begin{obsA} \label{obs:40} Given a linear logic context $\Delta_0 =
  [\alpha | \alpha = \llbracket s \rrbracket^{T}, s \in gc(G_0) ]$
  (types of the ground components of $G_0$), a multiset $\Gamma$
  including the referring typed variables for $\Delta_0$, and a
  multiset $\Gamma_P = [\rho | \rho = \llbracket \pi(p)
  \rrbracket^{T}, p \in P ]$ (types of the transformation rules), for
  every graph $G$ reachable in the system

$$\Gamma, \Gamma_P; \Delta_0 \Vdash \llbracket G \rrbracket^{T}$$

Given a multiset $R$ of transformations in $\mcal{G}$, let $\Delta_R =
[\tau | \tau = \llbracket t \rrbracket^{T}, t \in R ]$. Then, for each
graph $G$ which is reachable from $G_0$ by executing the
transformations in $R$, in some order

$$\Gamma; \Delta_0, \Delta_R \Vdash \llbracket G \rrbracket^{T}$$

If $G$ is reachable by executing at least the transformations in $R$,
in some order

$$\Gamma,\Gamma_P; \Delta_0, \Delta_R \Vdash \llbracket G \rrbracket^{T}$$
\end{obsA}

A further topic that we would like to investigate is concurrency. The
expressiveness of linear logic makes it comparatively natural to
represent parallel application of rules, choice and indeterminism, and
therefore to compare this embedding with classic graph transformation
approaches \cite{EEPT06}.


\section{Conclusion}

We have defined a translation of DPO GTS, formulated in algebraic
terms, with a restriction to rules that have only nodes in the
interface, into a quantified version of ILL, based on linear
$\lambda$-calculus, extended with a resource-bound existential
quantifier that we have used to type name restriction in graph
expressions. We have proved informally that the translation is sound
and complete with respect to graph expressions and adequate with
respect to reachability in GTS. We believe that a line of research
that relates models based on graph transformation and proof theory
along lines such as those of the \emph{Chemical Abstract Machine}
\cite{Berry90} is probably worth further investigation. Related work
on the translation of multiset rewriting into ILL has been discussed
for example in \cite{CeSc06}. We would like to mechanise the logic on
a theorem prover, and we are considering Isabelle, for which there is
already a theory of ILL \cite{Dixon06}.








\small
\bibliographystyle{eptcs}
\bibliography{biol,fagrat}

\end{document}